\documentclass{appolb}
\usepackage{graphicx}
\begin{document}
% \eqsec  % uncomment this line to get equations numbered by (sec.num)
\title{FINE STRUCTURE OF GIANT RESONANCES: \\ WHAT CAN BE LEARNED}%
%\thanks{Presented at the Zakopane Conference on Nuclear Physics “Extremes of the Nuclear Landscape”, Zakopane, Poland, August 26 -- September 2, 2018.}%
% you can use '\\' to break lines

\author{Peter von Neumann-Cosel
\address{Institut f\"ur Kernphysik, Technische Universit\"at Darmstadt, Germany}
}
\maketitle
\begin{abstract}
Fine structure of giant resonances (GR) has been established in recent years as  a global phenomenon across the nuclear chart and for different types of resonances. 
A quantitative description of the fine structure in terms of characteristic scales derived by wavelet techniques is discussed.
By comparison with microscpic calculations of GR strength distributions one can extract information on the role of different decay mechanisms contributing to the width of GRs. 
The observed cross-section fluctuations contain information on the level density (LD) of states with a given spin and parity defined by the multipolarity of the GR.
\end{abstract}

%\PACS{
%24.30.Cz 	Giant resonances;
%21.10.Ma 	Level density
%}
  
\section{Introduction}

Giant resonances are elementary excitations of the nucleus and their understanding forms a cornerstone of microscopic nuclear theory.
They are classified according to their quantum numbers (angular momentum, parity, isospin). 
Gross properties like energy centroid and strength in terms of exhaustion of sum rules are fairly well described by microscopic models \cite{har01}.
However, a systematic understanding of the decay widths is still lacking. 

\begin{figure}[htb]
\centerline{%
\includegraphics[width=8.5cm]{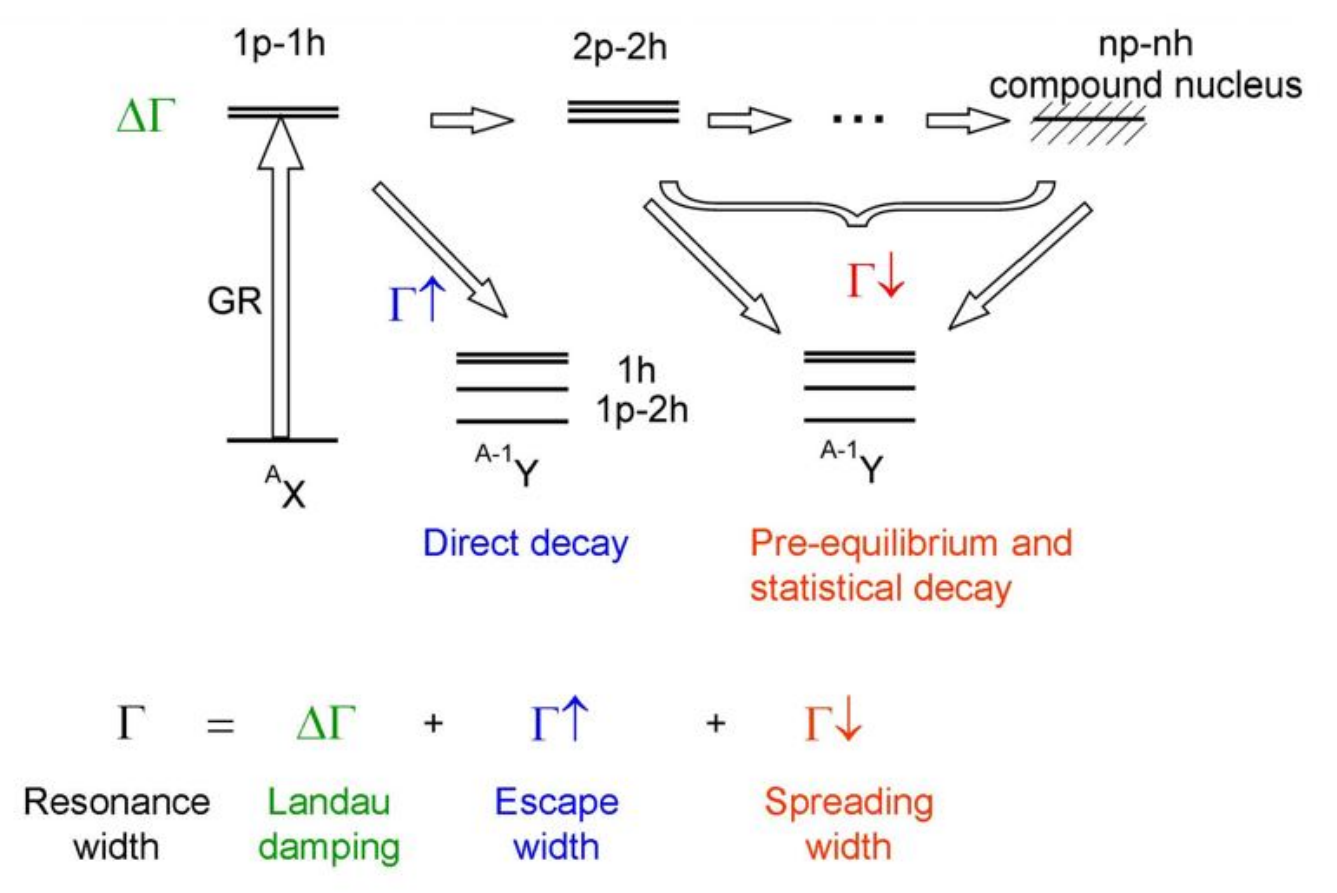}}
\caption{Schematic illustration of different decay mechanisms contributing to the width of giant resonances.}
\label{fig1}
\end{figure}

The giant resonance width $\Gamma$ is determined by the interplay of different mechanisms illustrated in Fig.~\ref{fig1}: fragmentation of the elementary one particle-one hole (1p-1h) excitations (Landau damping $\Delta E$)), direct particle decay out of the continuum (escape width $\Gamma\!\uparrow$), and statistical particle decay due to coupling to two (2p-2h) and many particle-many hole (np-nh) states (spreading width  $\Gamma\!\downarrow$) 
\begin{equation}
\label{eq:width}
\Gamma = \Delta E + \Gamma\!\uparrow + \Gamma\!\downarrow.
\end{equation}

A powerful approach to investigate the role of the different components are coincidence experiments, where direct decay can be identified by the population of one-hole states in the daughter nucleus and the spreading width contribution can be estimated by comparison with statistical model calculations (see, e.g., Refs.~\cite{bol88,die94,str00,hun03}).
Recently, an alternative method has been developed based on a quantitative analysis of the fine structure of giant resonances oberved in high-resolution inelastic scattering and charge-exchange reactions.
As demonstrated below, fine structure appears as a global feature of giant resonances across the nuclear chart. 
For comparable energy resolution, the fine structure properties are independent of the exciting probe \cite{kam97}, cf.\ upper left part of Fig.~\ref{fig2}.
Different approaches for an extraction of energy scales characterizing the phenomenon have been discussed in Ref.~\cite{she08}. 
Wavelet analysis has been identified as a particularly promising type of analysis.

In many cases, the cross section fluctuations are particularly pronounced on the low-energy side of the GRs and damped on the high-energy side.   
The magnitude of the fluctuations for a given experimental energy resolution is determined by the density of states, whose spin and parity is determined by the multipolarity of the GR.
If a single excitation mode dominates the cross sections and there is a way to estimate the background in the spectra, one can deduce the level density in the energy region of the GR with a fluctuation analysis.    

\section{Quantitative analysis of the fine structure}

\subsection{Experimental evidence for fine structure}

\begin{figure}[t]
\centerline{%
\includegraphics[width=12cm]{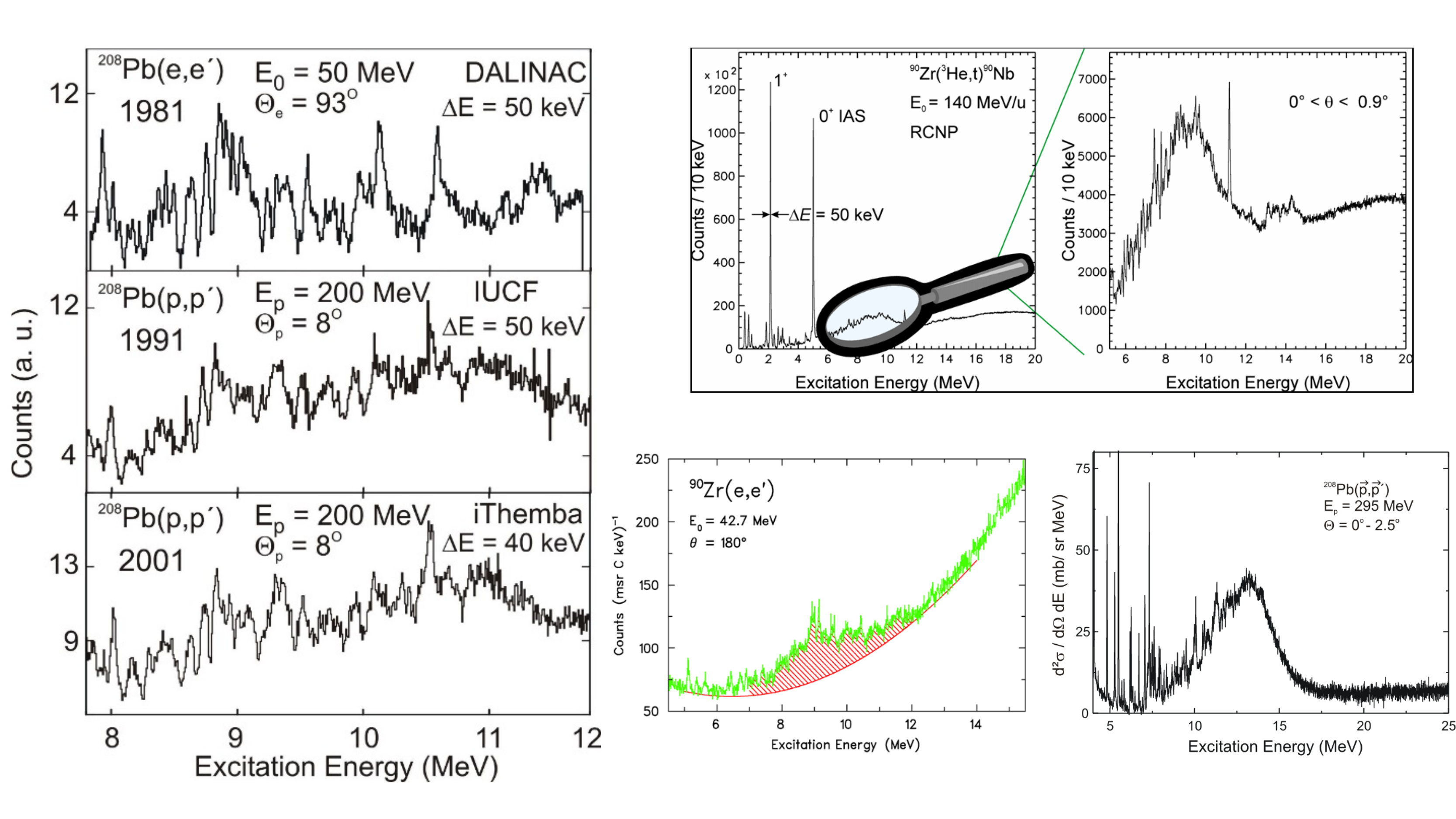}}
\caption{Examples of the fine structure phenomenon of different GRs in high-resolution experiments.
Left: ISGQR in $^{208}$Pb studied in (e,e$^\prime$) and (p,p$^\prime$) reactions \cite{kam97}.
Upper right: GTR in $^{90}$Nb studied with the $^{90}$Zr($^3$He,t) reaction at $0^\circ$ \cite{kal06}.
Lower middle: M2 resonance  in $^{90}$Zr studied in $180^\circ$ electron scattering \cite{vnc99}.
Lower right: IVGDR in $^{208}$Pb studied in the (p,p$^\prime$) reaction at $0^\circ$ \cite{tam11}. 
 }
\label{fig2}
\end{figure}

In recent years, systematic high-resolution  (p,p$^\prime$) experiments have been performed at iThemba LABS and RCNP to study the properties of the ISGQR \cite{she04,she09,usm11,usm17} and the IVGDR \cite{tam11,pol14,has15,fea18,jin18}, respectively.
Fine structure was observed across the nuclear chart.
It has also been demonstrated for M1 \cite{bir16}, M2 \cite{vnc99} and GT \cite{kal06} resonances. 
Some examples of such data are presented in Fig.~\ref{fig2}. 

\subsection{Wavelet analysis}

Wavelet analysis has been established as a tool to quantitatively analyze the fine structure of nuclear giant resonances.
It can be regarded as an extension of Fourier analysis which allows to conserve the correlation between the observable and its transform.
In the present application to nuclear spectra the coefficients of the wavelet transform are defined as
\begin{equation}
   C\left( {\delta E,E_{x}} \right) = \int\limits_{ - \infty
   }^\infty  {\sigma\left( E \right)\Psi \left( {\delta E,E_{x},E}
   \right)dE}.
   \label{eq:cwt}
\end{equation}
They depend on two parameters, a scale $\delta E$ stretching and compressing the wavelet $\Psi$($E$), and a position $E_{\rm x}$ shifting the wavelet in the spectrum $\sigma$($E$). 
The variation of the variables can be carried out with continuous or discrete steps. 
The analysis of the fine structure of giant resonances is performed using the continuous wavelet transform (CWT), where the fitting procedure can be adjusted to the required precision. 

In order to achieve an optimal representation of the signal using wavelet transformation, one has to select a wavelet function $\Psi$ which resembles the properties of the studied signal $\sigma$. 
A maximum of the wavelet coefficients at certain value $\delta$E indicates a correlation in the signal at the given scale, also called characteristic scale.
The best resolution for nuclear spectra is obtained with the so-called Complex Morlet wavelet (cf.\ Fig.~9 in Ref.~\cite{she08}), because the detector response closely resembles a Gaussian line shape. 
The Complex Morlet wavelet is a product of Gaussian and cosine functions
\begin{equation}
\Psi (x) = \frac{1}{\sqrt{\pi f_b}}{\rm exp}(2 \pi i f_c){\rm exp}\left(-\frac{x^2}{f_b}\right),
\label{eq:morlet}
\end{equation}
where $f_{\rm c}$ is the wavelength centre frequency and $f_{\rm b}$ is the bandwidth parameter. 

Alternatively, a spectrum decomposition based on the discrete wavelet transform (DWT) can be used, where scales and positions in the wavelet analysis are varied by powers of two. 
It allows an iterative decomposition of the spectrum by filtering it into two signals, approximations ($A$) and details ($D$), representing the large-scale (low-frequency) and small-scale (high-frequency) part for a given scale region analog to the effect of high- and low-pass filters in an electric circuit. 
In each step $i$ of the decomposition, the initial signal $\sigma$($E$) can be reconstructed as
\begin{equation}
\sigma(E)=A_i + \sum D_i.
\label{eq:dwtapp+det}
\end{equation}
This operation can be repeated until the individual detail consists of a single bin. 

A DWT can only be performed with wavelets which possess a so-called scaling function \cite{she08}.
This is not the case for the Complex Morlet wavelet, thus the Bior wavelet family \cite{ber99} is used as an alternative.  
It provides another useful property for a determination of background in the data, which is a prerequisite for the level density extraction described below. 
Each wavelet function can be characterized by its number of vanishing moments,
\begin{equation}
   \label{eq:vanishingm}
   \int\limits_{ - \infty }^\infty  {E^n \Psi \left( E \right)dE =
   0,\;\; n = 0,1...m}.
\end{equation}
Thus, any smooth background in the spectrum that can be approximated by a polynomial function up to order $m-1$ does not contribute to the wavelet coefficients. 
Examples are discussed in Refs.~\cite{she08,usm11,pol14,kal06}.
One can identifiy in all cases the decomposition order $i$ containing the largest scale, i.e.\ the resonance width.
The next-higher order provides the form of the background.

\subsection{Example: Characteristic scales of the ISGQR in $^{208}$Pb}

\begin{figure}[tbh]
\centerline{%
\includegraphics[height=6.75cm,angle=90]{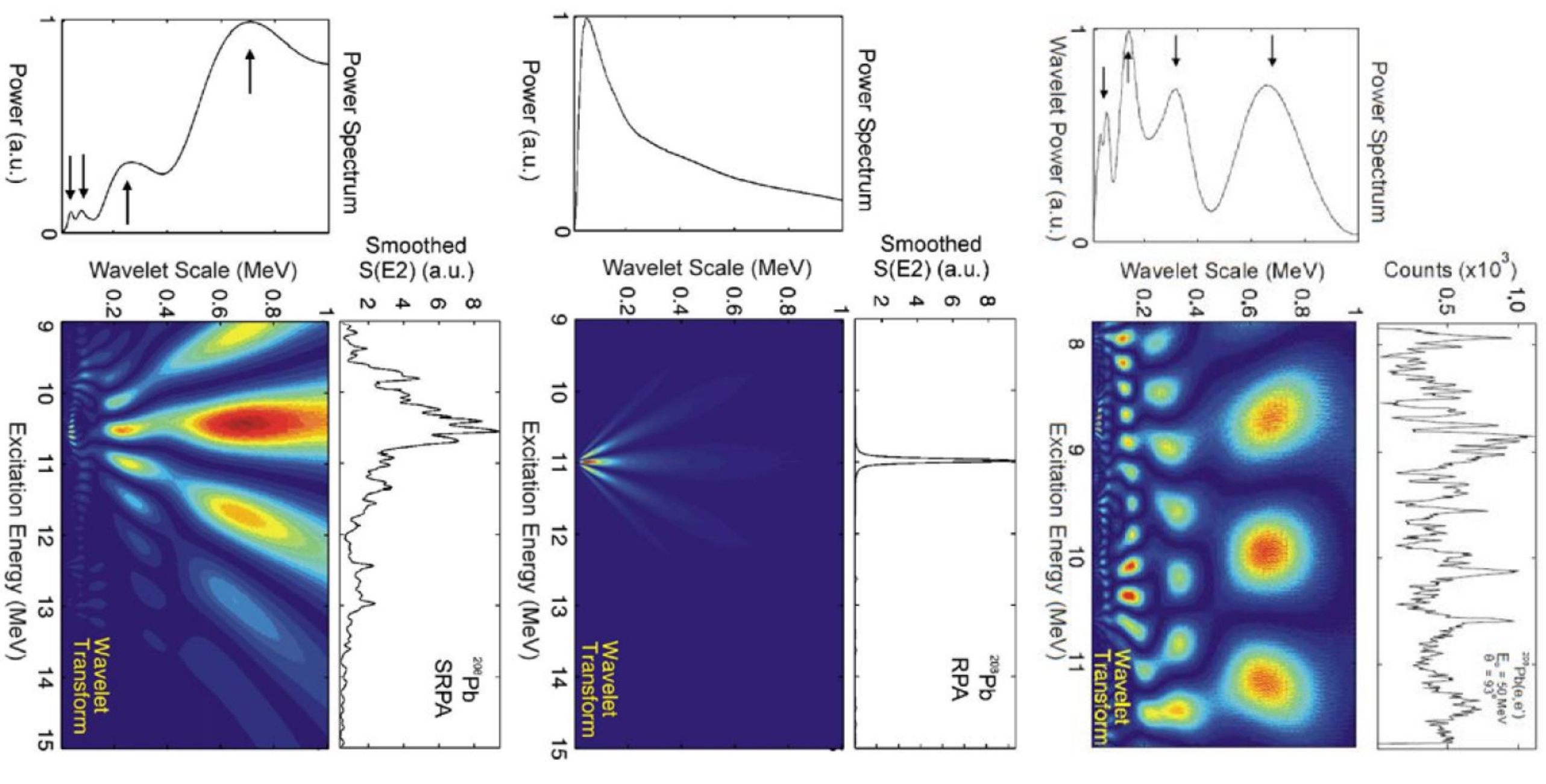}}
\caption{Top: Spectrum of the $^{208}$Pb (e,e$^\prime$) reaction \cite{kuh81}, squares of the wavelet coefficients as a function of excitation energy from a CWT,  and projection of the wavelet coefficients on the scale axis (power spectrum).
Middle: Same for a RPA calculation.
Bottom: Same for a SRPA calculation.}
\label{fig3}
\end{figure}

The exctraction of characteristic scales and their interpretation is discussed by way of example for a study of the ISGQR in $^{208}$Pb with the (e,e$^\prime$) reaction \cite{kuh81} (see l.h.s.\ of Fig.~\ref{fig3}). 
The 2D distribution of the squared wavelet coefficients of  the experimental spectrum show pronounced maxima at certain scale values across the energy region of the ISGQR.
Their values can be determined from the projection on the scale axis (the power spectrum).
The middle part of Fig.~\ref{fig3} displays the same type of analysis for a RPA calculation \cite{she09} of the ISGQR in $^{208}$Pb.
The strength is concentrated in a single peak and correspondingly the wavelet power spectrum does not show any scales (the maximum at small scale values results from folding of the strength distribution with the  experimental resolution).
If one includes 2p-2h states in a SRPA calculation \cite{she09}, fine structure in the strength distribution and corresponding maxima in the power spectrum are observed demonstrating that these characteristic scales arise from the damping width.
More specifically, the scales result from coupling to low-lying vibrations \cite{she04,she09}, a damping mechanism discussed in Ref.~\cite{ber83}.  

\subsection{Example: $K$ splitting of the ISGQR in deformed nuclei}

\begin{figure}[t]
\centerline{%
\includegraphics[width=13cm]{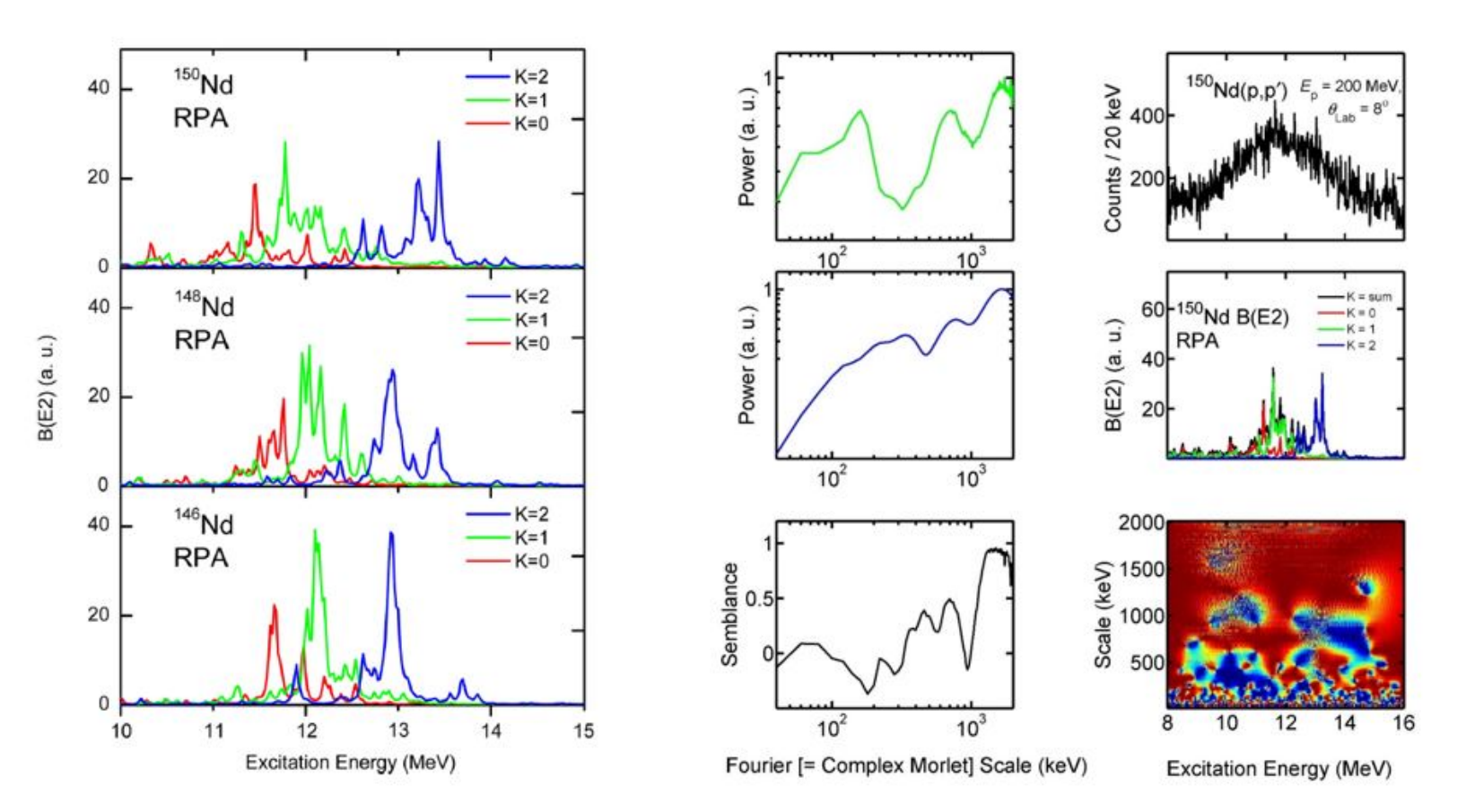}}
\caption{Left: RPA calculation of the splitting of the ISGQR in $^{146,148,150}$Nd into $K = 0,1,2$ components \cite{kur18}.
Right, top and middle row: Experimental and RPA strength distributions of $^{150}$Nd and wavelet power spectra.
Right, bottom row: Semblance analysis, Eq.~(\ref{eq:semblance}), and corresponding semblance power \cite{kur18}. 
} 
\label{fig4}
\end{figure}

The IVGDR in heavy deformed nuclei exhibits a characteristic double-hump structure identified as    splitting due to the conservation of the $K$ quantum number \cite{har01}. 
A similar splitting is predicted for the ISGQR as illustrated in the l.h.s.\ of Fig.~\ref{fig4} showing QRPA calculations of the ISGQR in the nuclei $^{146,148,150}$Nd with increasing deformation using the SVmas10 \cite{klu09} Skyrme interaction. 
The energy splitting is largest between the $K = 0$ and $K = 1,2$ components and increases with mass  number, but it is generally smaller than their typical widths. 
Therefore, $K$ splitting cannot be observed in a measurement of the ISGQR strength function.
However, it was shown recently that the fine structure may carry a signature \cite{kur18}.

The r.h.s.\ of Fig.~\ref{fig4} illustrates an application of the CWT on both experimental and theoretical spectra of the ISGQR in $^{150}$Nd. 
The top and middle row show the experimental and theoretical spectrum, respectively, and the wavelet power spectra, derived as described above.
Although no scales are observed on the theoretical power spectrum for smaller scale values, pronounced characteristic scales are visible around 1 MeV resembling the experimental results. 
These are caused by the splitting between the main fragments of $K$ = 0, 1 and 2 strengths.

This interpretation can be further tested by a semblance analysis, which provides a quantitative measure of the correspondence between two sets of wavelet coefficents by studying the local phase relationships of the complex wavelet coefficients as a function of scale \cite{coo08}.
The semblance $S$ can be expressed as
\begin{equation}
\label{eq:semblance}
S = \cos^n (\theta),
\end{equation}
where $n$ is an odd integer greater than zero ($n = 1$ in the present case), yielding values ranging from -1 (anticorrelated) through 0 (uncorrelated) to +1 (correlated).   
Here, the local phase $\theta$ is given by $\theta = \tan^{-1} [\Im(C_{1,2})/\Re(C_{1,2})]$, where the cross-coefficient $C_{1,2} = C_1 C^*_2$ with $C_1$ the wavelet transform of data set 1 and $C^*_2$ the complex conjugate of dataset 2. 

The bottom row shows the result from the application of Eq.~(\ref{eq:semblance}) to the experimental spectrum and the RPA prediction.
For smaller scale values the semblance shows large fluctuations from correlation (red) to anti-correlation (blue) over the energy region of the resonance.
A large positive correlation is obtained over most of the resonance  -- in this case between $E_{\rm x}$ = 11 to 13 MeV where the RPA $E2$ strength lies -- for scale values corresponding to two characteristic scales around 1 MeV supporting the relation of these power maxima to the $K$ splitting.
Further details can be found in Ref.~\cite{kur18}.

\section{Level densities}

The magnitude of fluctuations of cross section observed in high-resolution experiments in the energy region of the GRs is related to the LD.
It can be extracted with a fluctuation analysis decribed, e.g., in Refs.~\cite{pol14,kal06,usm11a,mar17}.

\begin{figure}[t]
\centerline{%
\includegraphics[width=12cm]{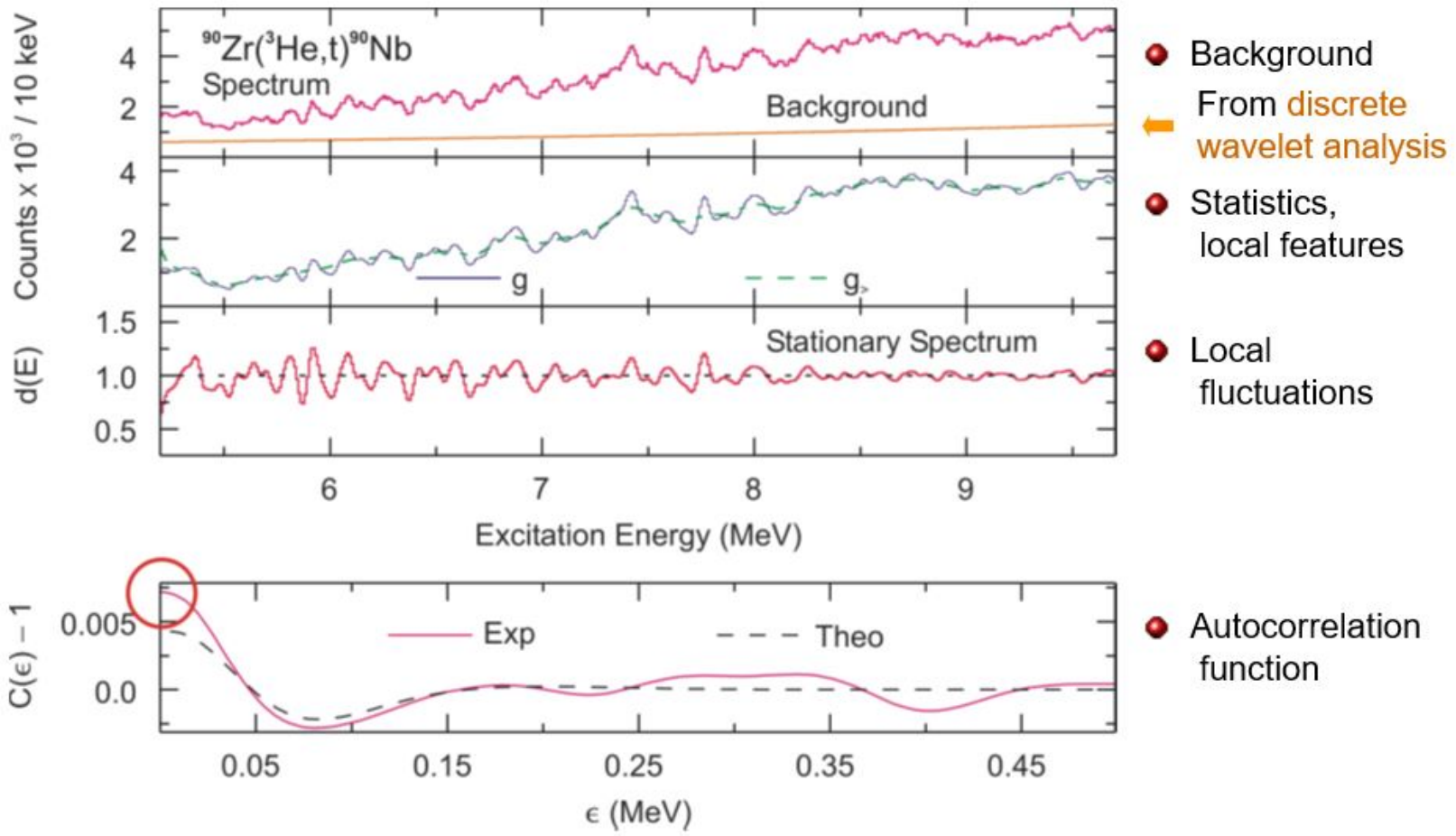}}
\caption{Top row: Spectrum of the $^{90}$Zr($^3$He,t) reaction at $0^\circ$ \cite{kal06}
and background obtained from a DWT (orange line). 
Second row: Background-subtracted smoothed spectra $g(E_x)$ and $g_>(E_x)$. 
Third row: Stationary spectrum $d(E_x)$. 
Bottom row: Experimental [Eq.~(\ref{eq:autocorexp})] and theoretical [Eq.~(\ref{eq:autocorrtheo})] autocorrelation function .}
\label{fig5}
\end{figure}

The procedure of the fluctuation analysis is schematically demonstrated in Fig.~\ref{fig5} for the example of the Gamow-Teller GR in $^{90}$Nb measured with the $^{90}$Zr($^3$He,t) reaction \cite{kal06}.
It can be divided in four main steps.
The corresponding spectrum in the region of interest (cf.\ Fig.~\ref{fig2}) is shown in the top row of Fig.~\ref{fig5}.
For an extraction of the LD one has to subtract any background not arising from excitations of the nuclear mode under investigation. 
In the present example it was determined by a DWT analysis as described in Ref.~\cite{kal06}.

Further, fluctuation contributions arising from finite statistics are removed by folding with a Gaussian function of width $\sigma$ chosen to be smaller than the experimental energy resolution. 
The resulting spectrum is called $g(E_x)$ hereafter. 
Similarly, a second spectrum $g_>(E_x)$ is created by the convolution with a Gaussian function, whose width $\sigma_>$ is at least two times larger than the energy resolution in the experiment in order to remove gross structures from the spectrum.
The spectra $g(E_x)$ and $g_>(E_x)$ for the present data are shown in the second row of Fig.~\ref{fig5}.
The dimensionless stationary spectrum $d(E_x) = g_>(E_x )/g(E_x)$
%\begin{equation}
%    \label{eq:stsection}
%       d\left( {E_x } \right) =
%       \frac{{g_> \left( {E_x } \right)}}{{g \left( {E_x } \right)}}
%\end{equation}
is shown in the third row.
As a result of the normalization on the local mean value, the energy dependence of the cross sections vanishes. 
The value of $d(E_x)$ is sensitive to the fine structure of the spectrum and distributed around an average intensity $\langle d(E_x)\rangle$=1. 
With increasing excitation energy the mean level spacing is decreasing, and in turn the oscillations of $d(E_x)$ are damped.

A quantitative description of the fluctuations is given by the autocorrelation function
\begin{equation}
    \label{eq:autocorexp}
        C\left( \epsilon  \right) =
        \frac{{\left\langle {d\left( {E_x } \right) \cdot d\left( {E_x  +
        \epsilon } \right)} \right\rangle }}{{\left\langle {d\left( {E_x }
        \right)} \right\rangle  \cdot \left\langle {d\left( {E_x  +
        \epsilon } \right)} \right\rangle
        }}\;.
\end{equation}
The value $C(\epsilon = 0) - 1$ is nothing but the variance of $d(E_x)$
\begin{equation}
    \label{eq:autocorvar}
        C\left( {\epsilon  = 0} \right) - 1 = \frac{{\left\langle {d^2
        \left( {E_x } \right)} \right\rangle  - \left\langle {d\left( {E_x
        } \right)} \right\rangle ^2 }}{{\left\langle {d\left( {E_x }
        \right)} \right\rangle ^2 }}\;.
\end{equation}
According to Ref.~\cite{jon76}, this experimental autocorrelation function shown in the bottom row of Fig.~\ref{fig5} can be approximated by the expression
\begin{equation}
\label{eq:autocorrtheo}
C(\epsilon) - 1 =  \frac{\alpha \cdot \langle \mbox{D} \rangle}{2 \Delta E \sqrt{\pi}} \times f(\sigma,\sigma_>),
\end{equation}
where the function $f$ depends on the chosen parameters (folding widths $\sigma, \sigma_>$) only.
The value $\alpha$ is the sum of the normalized variances of the assumed spacing and transition width distributions.
If only transitions with the same quantum numbers ($J^\pi =1^+$ in the present case) contribute to the spectrum, then $\alpha$ can be directly determined as the sum of the variances of the Wigner and Porter-Thomas distribution and the mean level spacing $\langle D\rangle$ can be extracted from the value of $C(\epsilon = 0) - 1$. 
The corresponding LD is given by $\rho(E)=1/\langle D\rangle$.

\begin{figure}[tbh]
\centerline{%
\includegraphics[width=8cm]{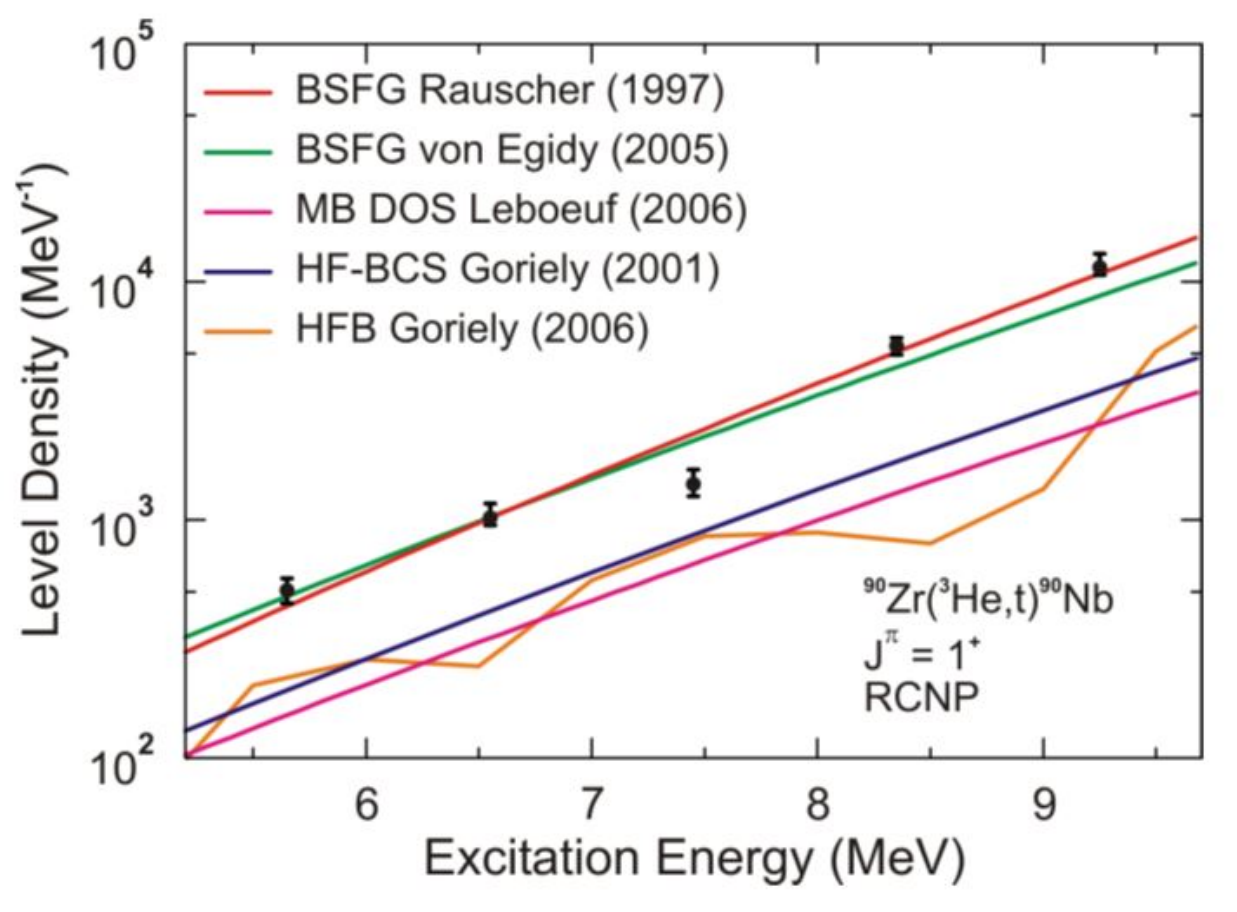}}
\caption{LD of $1^+$ states in $^{90}$Nb extracted with the fluctuation analysis from the data of Fig.~\ref{fig5} and comparison with different models (see text).}
\label{fig6}
\end{figure}

Figure \ref{fig6} depicts the results of the procedure for the $^{90}$Zr($^3$He,t) data for excitation energies of about 5 to 9 MeV in comparison with a variety of models.
Empirical parameterizations \cite{rau97,egi05} with the backshifted Fermi gas model (BSFG) describe the data well.
Microscopic calculations in the HF-BCS \cite{dem01} and HFB \cite{gor08} frameworks as well as a two-component Fermi gas (MB-DOS) \cite{leb06} underpredict the absolute density of $1^+$ states.

\section{Concluding remarks}

Systematic studies with light-ion induced reactions and electron scattering utilizing high-resolution spectrometers have demonstrated that fine structure of giant resonances is a global phenomenon. 
The present contribution discusses ways to quantitatively extract information from the observed fine structure and wavelet analysis has been established as the most promising tool.
It allows the extraction of scales in the power spectrum, which can be related to different decay mechanisms contributing to the width of GRs.
While this can also be achieved with a Fourier transform (and even with somewhat better resolution), the information from the wavelet transform, Eq.~(\ref{eq:cwt}), is essential to relate the origin of scales to the GRs.   
 
In kinematics where a particular GR dominates the spectra, one can extract LDs from the cross-section fluctuations by an autocorrelation analysis.
These LD results are quite unique in several aspects: 
(i) One obtains LD values for a specific spin and parity.
(ii) The method provides absolute values in contrast to LDs from the two major sources of LD data besides neutron resonance spacings, viz.\ the Oslo method \cite{sch00} and particle emission spectra \cite{voi14}.   
(iii) LD data above the particle thresholds are rare.  
(iv) They contribute to the resolution of important open questions like a possible parity dependence in certain shell regions \cite{kal07}, collective enhancement factors describing the role of vibrations and rotations in deformed nuclei \cite{oze13}, or the spin distribution of a given total LD.
The latter can be addressed by extracting LDs of $J =0,1,2$ states from the corresponding GRs (ISGMR, IVGDR, ISGQR, M1, M2) in the same nucleus.

I thank the many colleagues who contributed to the experiments at the S-DALINAC, iThemba LABS and RCNP.
I am particularly greatful to R.~Fearick, A.~Richter and J.~Wambach for their important contributions to the development of the described methods and countless discussions. 
This work has been supported by the DFG under contract SFB 1245.

\end{document}